\documentclass{jpsj-suppl}
\usepackage{times}
\title{Analysis of Magnetic Field-Angle Dependent Electronic Raman Scattering to Probe the Superconducting Gap}

\author{
Masaru Okada$^{1}$\thanks{E-mail address: o.masaru@issp.u-tokyo.ac.jp}, \
Nobuhiko Hayashi$^{2}$
}
\inst{
$^{1}$ Institute for Solid State Physics, University of Tokyo, 5-1-5 Kashiwanoha, Kashiwa, Chiba 277-8581, Japan \\
$^{2}$ Nanoscience and Nanotechnology Research Center (N2RC), Osaka Prefecture University, 1-2 Gakuen-cho, Naka-ku, Sakai 599-8570, Japan
}
\abst{
(Received September 30, 2013)
\\ \\
We study the field-angle resolved electronic Raman scattering in 2-dimensional $d$-wave superconducting vortex states theoretically by quasi-classical approximation, the so-called Doppler-shift method.
An analytic expression is obtained for the field-angle dependence of the Raman scattering amplitude at zero temperature.
After numerical integration, we obtain the electronic Raman scattering intensity for various field angles by changing the Raman shift energy.
Field-angle resolved electronic Raman scattering turns out to be an effective method for probing unconventional superconducting gap structures.
It shows a novel phenomenon: reversal of extrema as a function of frequency without changing temperature or field magnitude.
}

\kword{
superconductivity,
vortex lattice,
electronic Raman scattering,
quasi-classical approximation,
Doppler shift
}

\begin{document}
\maketitle

\def\vector#1{\mbox{\boldmath $#1$}}


\section{Introduction}

Some kinds of unconventional superconductors have anisotropic order parameters,
such as the high-$T_{c}$ superconducting cuprates, which have $d$-wave symmetry
\cite{Cooper1989}.
Determining the symmetry of the order parameter is essential
for understanding the pairing mechanism.

\begin{figure}[b]
\begin{center}
\includegraphics[width=9cm]{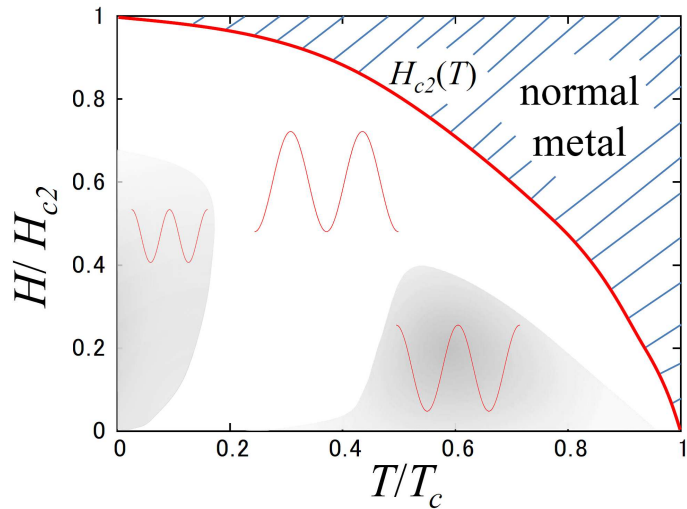}
\end{center}
\caption{
The field-temperature phase diagram
which shows that extrema in the symmetry patterns are reversed among some regions.
This schematic diagram is redrawn from Fig.12 in Ref. \cite{VorontsovPRB2007}.
}
\label{f1}
\end{figure}

The field-angle dependent specific heat and thermal conductivity experiments have been
studied to probe superconducting pairing symmetry in novel materials,
such as a family of heavy-fermion compounds Ce$M$In$_{5}$ ({\it M} = Rh, Co, and Ir) \cite{Izawa2001,KasaharaPRL2008,An2010}.
Fig. \ref{f1} shows the field-temperature phase diagram for the normalized fourfold thermal conductivity $\kappa_{4 \alpha}$
in magnetic field $\vector{H}$; $\vector{H}$ is rotated within the $a$-$b$ plane at angle $\alpha$.
The shaded (unshaded) regions correspond to minima (maxima) of $\kappa_{4 \alpha}$ for $\vector{H}$ along the nodal directions
\cite{VorontsovPRB2007}.
In this figure, minima and maxima are reversed upon changing temperature or field magnitude across critical boundaries.
This reversal means that gap symmetry cannot be determined by this experiment alone.

The electronic Raman scattering experiment can provide the information in both real and momentum space.
In addition, energy (frequency) is the essential variable for the electronic Raman scattering experiment.
Through this experiment, we can measure excitations from low energy to high energy.
Field-angle resolved electronic Raman scattering should be a useful technique for studying the order parameter.

%
%
%
%
%
%
%
%


\section{Formalism}

This section briefly summarizes
how to describe the electronic Raman scattering theoretically
\cite{Klein1984,Monien1990,Devereaux1995,Jiang1996,Devereaux2007}.
We have unit quantities $\hbar$, $k_{\rm B}$, $c$, and $e$.

The electronic Raman scattering amplitude is proportional to the imaginary part of the response function
%
%
%
%
\begin{eqnarray}
{\rm Im}
\chi^{(R)}_{\tilde{\rho} \tilde{\rho}} (\vector{q},\omega)
&=&
{\rm Im} \ \!
\Big[ i \int^{\infty}_{0} \!\!\! dt \ \! \big\langle [ \tilde{\rho}_{\vector{q}}(t) , \tilde{\rho}^{\dagger}_{\vector{q}} ] \big\rangle \ \! e^{i \omega t} \Big]
,
\end{eqnarray}
%
%
%
%
where $\tilde{\rho}_{\vector{q}}$ is the effective Raman operator, which is defined as
%
%
%
%
\begin{eqnarray}
\tilde{\rho}^{\dagger}_{\vector{q}}
&=&
\sum_{\vector{k},\sigma}
\gamma_{\vector{k}}
c_{\vector{k}+\vector{q},\sigma}^{\dagger} c_{\vector{k},\sigma}
,
\end{eqnarray}
%
%
%
%
with $c_{\vector{k},\sigma}$ being the annihilation operator of the electron with momentum $\vector{k}$ and spin $\sigma$.
The electronic Raman scattering experiment yields information on the interaction
between the electron and the photon.
This interaction is given by the perturbative Hamiltonian
%
%
%
%
\begin{eqnarray}
H'
&=&
\frac{1}{2m}\sum_{j}
\left[ \vector{p}_{j}
+\vector{A}_{\rm I}(\vector{r}_{j})
+\vector{A}_{\rm S}(\vector{r}_{j})
\right]^{2}
-
\frac{1}{2m}\sum_{j}
{\vector{p}_{j}}^{\ 2}
\nonumber \\[3mm] &=&
{H'}^{(1)} + {H'}^{(2)} + {H'}^{(A)}
,
\end{eqnarray}
%
%
%
%
where
%
%
%
%
\begin{eqnarray}
\left\{
\begin{array}{rcl}
H'^{(1)}
&=&
\frac{1}{2m}\sum_{j}
2 \left[ \vector{p}_{j} \cdot  \left\{ \vector{A}_{\rm I}(\vector{r}_{j}) + \vector{A}_{\rm S}(\vector{r}_{j}) \right\} \right]
\\[3mm]
H'^{(2)}
&=&
\frac{1}{2m}\sum_{j}
2 \vector{A}_{\rm I}(\vector{r}_{j}) \cdot \vector{A}_{\rm S}(\vector{r}_{j})
\end{array}
\right.
.
\end{eqnarray}
%
%
%
%
The vector potentials at position $\vector{r}_{j}$ are
%
%
%
%
\begin{eqnarray}
\left\{
\begin{array}{rcl}
\vector{A}_{\rm I}( \vector{r}_{j} )
&=&
\vector{\hat{e}}_{\rm I} \frac{A_{\rm I}}{2}
\big( {a_{\rm I}}^{\dagger} e^{i {\vector{K}}_{\rm I} \cdot \vector{r}_{j} }+ a_{\rm I} e^{-i {\vector{K}}_{\rm I} \cdot \vector{r}_{j} } \big)
,
\\[2mm]
\vector{A}_{\rm S}( \vector{r}_{j} )
&=&
\vector{\hat{e}}_{\rm S} \frac{A_{\rm S}}{2}
\big( {a_{\rm S}}^{\dagger} e^{i {\vector{K}}_{\rm S} \cdot \vector{r}_{j} }+ a_{\rm S} e^{-i {\vector{K}}_{\rm S} \cdot \vector{r}_{j} } \big)
\end{array}
\right.
,
\end{eqnarray}
%
%
%
%
where $a_{\rm I}$ ($a_{\rm S}$) is the annihilation operator of the incident (scattering) photon and
$\vector{\hat{e}}_{\rm I}$ ($\vector{\hat{e}}_{\rm S}$) is the unit vector along the polarization of incident (scattered) light.
The transfer momentum $\vector{q}$ can be found by $\vector{q}=\vector{K}_{\rm S} - \vector{K}_{\rm I}$.
The Raman vertex $\gamma_{\vector{k}}$ is defined by the standard calculation of the 2nd order perturbation theory,
%
%
%
%
\begin{eqnarray}
\langle f | \gamma_{\vector{k}} c_{\vector{k}+\vector{q},\sigma}^{\dagger} c_{\vector{k},\sigma} | i \rangle
&=&
\langle f | H'^{(2)} | i \rangle
+
\sum_{ \{ m | \varepsilon_{m} \neq \varepsilon_{\vector{k}} \} }
\dfrac{ \langle f | H'^{(1)} | m \rangle \langle m | H'^{(1)} | i \rangle  }{ \varepsilon_{\vector{k}} - \varepsilon_{m} }
,
\end{eqnarray}
%
%
%
%
where
$| i \rangle$ and $| f \rangle$ stand for the initial and final state.
After some calculation, we find that $\gamma_{\vector{k}}$ can be written in terms of the curvature of the energy band dispersion $\varepsilon_{\vector{k}}$ as
%
%
%
%
\begin{eqnarray}
\gamma_{\vector{k}}
&=&
\dfrac{ \partial^{2} \varepsilon_{\vector{k}} }
{ \partial (\vector{k} \cdot \vector{\hat{e}}_{\rm I}) \partial (\vector{k} \cdot \vector{\hat{e}}_{\rm S}) }
.
\end{eqnarray}
%
%
%
%

%
%
%
%
\begin{figure}[b]
\begin{center}
\includegraphics[width=7cm]{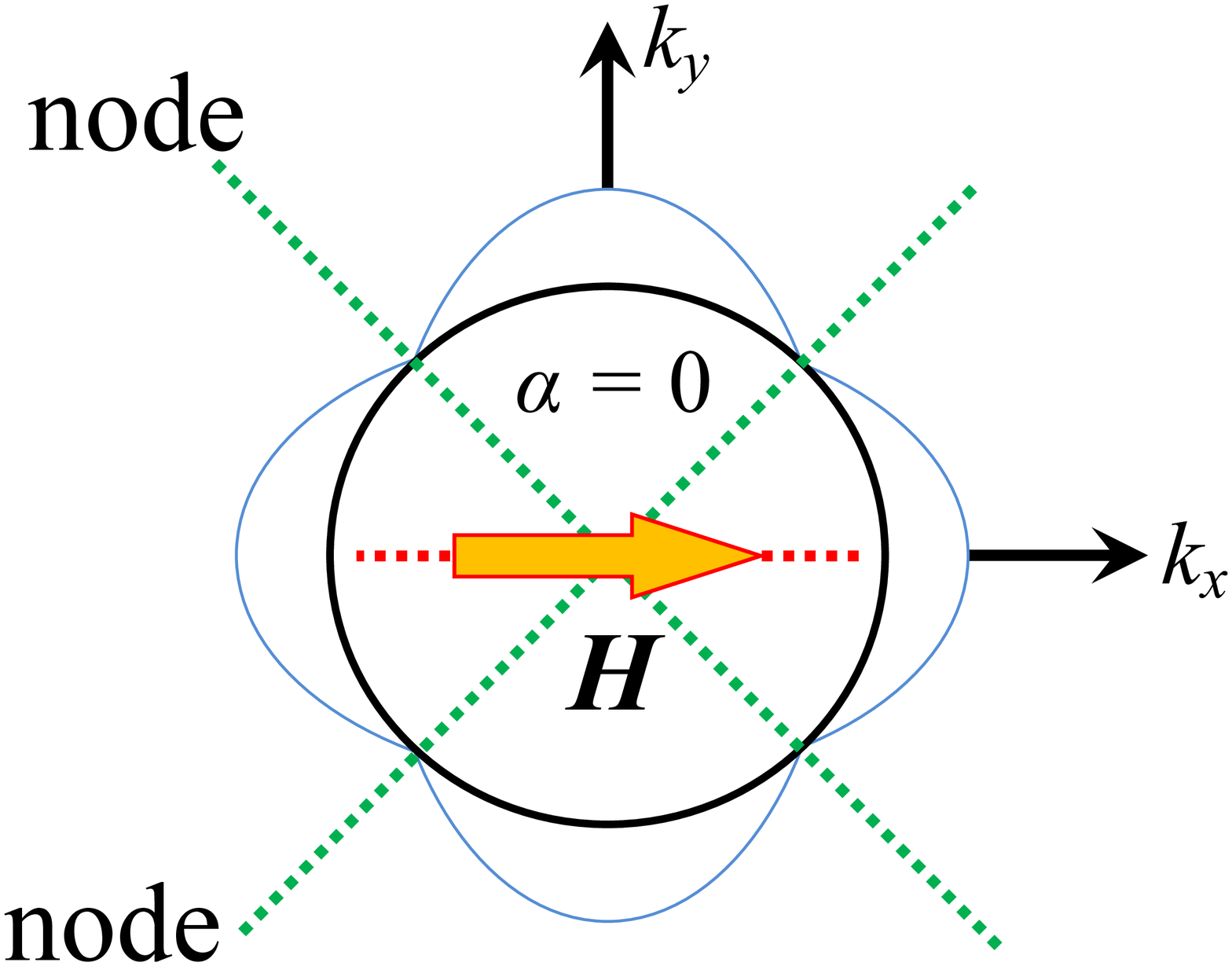}
\includegraphics[width=7cm]{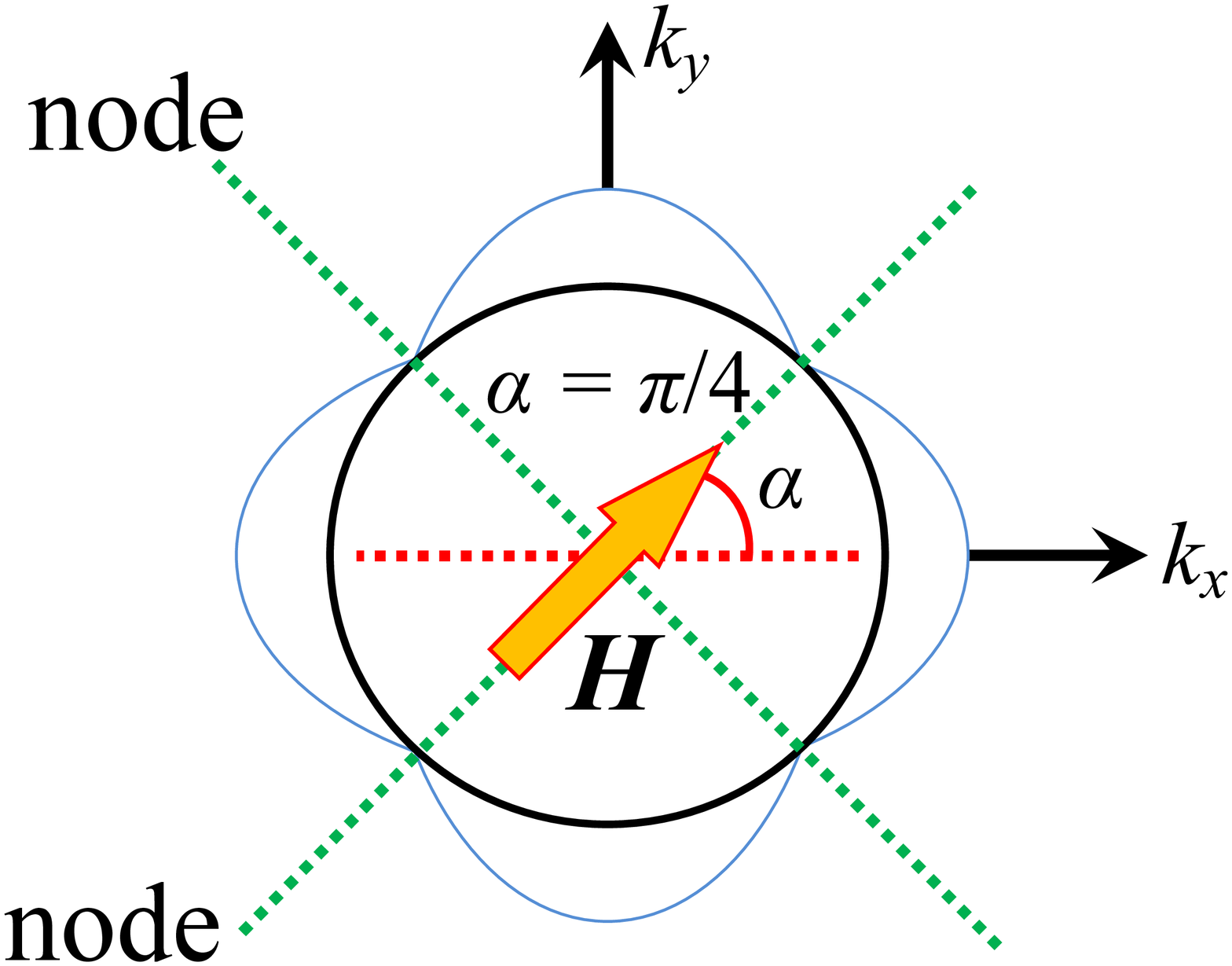}
\end{center}
\caption{
The Fermi surface with
the magnetic field $\vector{H}$ in the $a$-$b$ plane, at angle $\alpha=0,\pi/4$.
}
\label{f2}
\end{figure}
%
%
%
%

We consider a magnetic field $\vector{H}$ applied in the $a$-$b$ plane
at an angle $\alpha$
and account for its effect on the quasiparticle states by the Doppler energy shift
\cite{Vekhter1999,Dahm2002,Vekhter1999-2}
%
%
%
%
\begin{eqnarray}
\delta \omega_{\vector{k}}(\vector{r})
&=&
\vector{v}_{s} \cdot \vector{k}
\ = \
\dfrac{E_{H}}{(r/R)} {\rm sin} \beta {\rm sin} (\phi - \alpha)
,
\end{eqnarray}
%
%
%
%
where
$E_{H}=a \Delta_{0} \sqrt{H/H_{c2}}$
is the energy scale associated with the Doppler shift,
$a$ is a constant of order unity,
the field magnitude $H$ is valid for $0 < E_{H}/\Delta_{0} \ll 1$,
$(r,\beta)$ gives the position vector $\vector{r}$ in polar coordinates,
and superfluid velocity $\vector{v}_{s} =\vector{\hat{\theta}}/2mr$
is approximated by the flow field of an isolated vortex.
Here, $\vector{\hat{\theta}}$ is a unit vector along the supercurrent
and $r$ is the distance from the center of the vortex.
Assuming a 2-dimensional cylindrical Fermi surface,
local quantities $f(\vector{r})$ have to be averaged over the unit cell of the vortex lattice,
which is approximated by a circle of radius $R$.

In Fig. \ref{f2},
we show the orientation of the magnetic field in
relation to the node structure of the superconducting order parameter of $d_{x^{2}-y^{2}}$ symmetry.
This symmetry is represented by $\Delta_{\vector{k}}=\Delta_{0} \Delta_{\phi} = \Delta_{0} {\rm cos} 2 \phi$
where $\phi$ is the angle between the $\vector{k}$ vector and the $a$-axis (or $k_{x}$-axis).
For example, when the magnetic field $\vector{H}$ is applied tangent to the $a$-$b$ plane in the antinodal direction ($\alpha=0$),
all four nodes contribute equally to the density of states (left panel).
However, when $\vector{H}$ is applied in the nodal direction ($\alpha=\pi/4$),
the effect of the Doppler shift on the density of states vanishes at $\phi = \pi / 4 , 5 \pi /4$ (right panel).

To obtain the expression of the response function with the effect of the Doppler shift,
we employ the one-particle Green's function; this is obtained by
introducing the Doppler shift into the BCS-Gorkov function
\cite{Kubert1998}
%
%
%
%
\begin{eqnarray}
\check{G}(\vector{k}, i \omega_{n};\vector{r})
&=&
- \dfrac{ ( i \omega_{n} - \vector{v}_{s} \cdot \vector{k}) \check{\tau}_{0} + \Delta_{\vector{k}} \check{\tau}_{1} + \xi_{\vector{k}} \check{\tau}_{3} }
{( \omega_{n} + i \vector{v}_{s} \cdot \vector{k} )^{2} + \xi_{\vector{k}}^{2} + \Delta_{\vector{k}}^{2}  }
,
\end{eqnarray}
%
%
%
%
where $\omega_{n} = (2n+1)\pi T$ is the fermionic Matsubara frequency,
$T$ is the temperature,
$\xi_{\vector{k}}$ is the energy of a quasiparticle with momentum $\vector{k}$
(measured with respect to the Fermi level),
and the $\check{\tau}_{i}$ are Pauli matrices.
Then, the response function,
without vertex corrections in the limit of zero transfer momentum,
can be written as
%
%
%
%
\begin{eqnarray}
-\big\langle
\chi_{\tilde{\rho} \tilde{\rho}} (i \nu_{n})
\big\rangle_{H(\alpha)}
&=&
\sum_{\vector{k},i \omega_{n}} \!\!
\int \!\! d^{2} \vector{r}
\ {\rm Tr}\Big[ \gamma_{\vector{k}}^{2} \check{\tau}_{3} \check{G}(\vector{k},i \omega_{n};\vector{r}) \check{\tau}_{3} \check{G}(\vector{k},i \omega_{n} - i \nu_{n} ;\vector{r}) \Big]
,
\end{eqnarray}
%
%
%
%
where
$\nu_{n} = 2 n \pi T$
is the bosonic Matsubara frequency.

Summing over the Matsubara frequency
and averaging over the unit cell of vortex lattice
$\int d^{2} \vector{r} (\cdots) = \int^{1}_{0} d \rho \int^{2 \pi}_{0} d \beta \rho (\cdots)$,
where $\beta$ is the vortex winding angle,
can be done analytically.
For the 2-dimensional Fermi surface,
summing over the momentum and averaging over the Fermi surface
are equivalent.
After analytic continuation to real frequencies,
$ i \nu_{n} \to \omega + i0 $,
we obtain the following expression
\cite{Vekhter1999-2},
%
%
%
%
\begin{eqnarray}
&&
- \big\langle {\rm Im} \chi_{\tilde{\rho} \tilde{\rho}}(x) \big\rangle_{H(\alpha)}
\nonumber \\[3mm] &\propto&
\int_{\rm FS}
\! \! \!
d \phi
\ {\rm Re} \! \left[
 \frac{ \gamma_{\phi}^{2} \Delta_{\phi - \alpha}^{2} }{ x \sqrt{ x^{2} - \Delta_{\phi - \alpha}^{2} } } \right]
\nonumber \\ && \hspace{2mm} \times
\int^{1}_{0}
\!\!
d \rho
\int^{2 \pi}_{0}
\!\!
d \beta
\ \rho
\bigg[
{\rm tanh}
\bigg(
\frac{ \Delta_{0} }{ 2T }
x
+
\frac{E_{H}}{\rho}
\frac{
{\rm sin}\beta
\ \!
{\rm sin} \phi
}{2T}
\bigg)
+
{\rm tanh}
\bigg(
\frac{ \Delta_{0} }{ 2T }
x
-
\frac{E_{H}}{\rho}
\frac{
{\rm sin} \beta
\ \!
{\rm sin} \phi
}{2T}
\bigg)
\bigg]
\nonumber \\[3mm] &\longrightarrow&
\int_{\rm FS}
\! \! \!
d \phi
\ {\rm Re} \! \left[
 \frac{ \gamma_{\phi}^{2} \Delta_{\phi - \alpha}^{2} }{ x \sqrt{ x^{2} - \Delta_{\phi - \alpha}^{2} } } \right]
\nonumber \\ && \hspace{2mm} \times
\left\{
\begin{array}{ll}
\pi - \dfrac{\pi}{2} w^{2}
& \hspace{5mm} (0 < w < 1)
\\
\left( 2 - w^{2} \right)
{\rm arcsin} \dfrac{1}{w} + \sqrt{ w^{2} -1}
& \hspace{5mm} ( w > 1)
\end{array}
\right.
\hspace{10mm} (T \to 0)
,
\label{eqn:e1}
\end{eqnarray}
%
%
%
%
where
$x=\omega/2 \Delta_{0}$ is the Raman shift energy normalized by the gap amplitude $\Delta_{0}$;
we write $w = E_{H} {\rm sin} \phi / (\Delta_{0} x)$ and $\rho = r/R$ for simplicity.
$\Delta_{\phi} = \Delta_{\vector{k}}/\Delta_{0}$ and $\gamma_{\phi}=\gamma_{\vector{k}}/\gamma_{0}$
are the normalized pairing and Raman vertex functions.


\section{Results and Discussion}

We consider the $d_{x^{2} - y^{2}}$ order parameter
$
\Delta_{\vector{k}} = \Delta_{0} {\rm cos} 2 \phi
$.
For this order parameter, $B_{1g}$ polarization shows the most characteristic frequency dependence of electronic Raman scattering in zero magnetic field.
The $B_{1g}$ symmetry is an irreducible representation of the $D_{4h}$ point group
(other examples of irreducible representations are $A_{1g}$, $B_{2g}$, and $E_{g}$).
Different types of electronic Raman excitations can be observed under different polarizations.
We calculate
$
-\big\langle
\chi_{\tilde{\rho} \tilde{\rho}} (x)
\big\rangle_{H(\alpha)}
$
at zero temperature
by using the Raman vertex
$
\gamma_{\vector{k}} = \gamma_{0}^{B_{1g}} {\rm cos} 2 \phi
,
$
for the $B_{1g}$ polarization,
where $\gamma_{0}^{B_{1g}}$ is a constant
and the field amplitude is given by $E_{H} = 0.1 \Delta_{0}$.

%
%
%
%
\begin{figure}[htbp]
\begin{center}
\includegraphics[width=15cm]{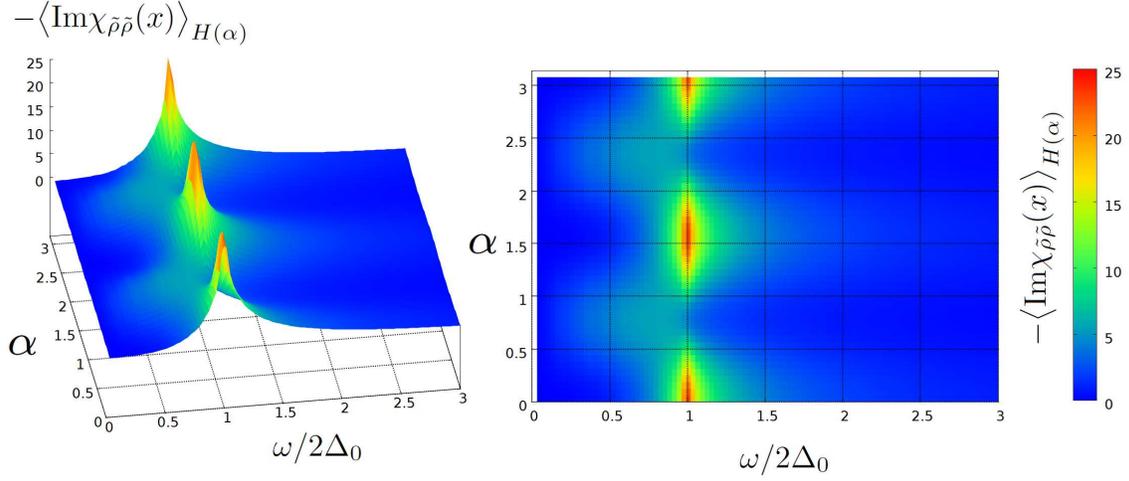}
\end{center}
\caption{
The field-angle dependent electronic Raman scattering intensity for various Raman energy shift $\omega$ with changing field angle $\alpha$ from 0 to $\pi$.
}
\label{f3}
\end{figure}
%
%
%
%

%
%
%
%
\begin{figure}[htbp]
\begin{center}
\includegraphics[width=7cm]{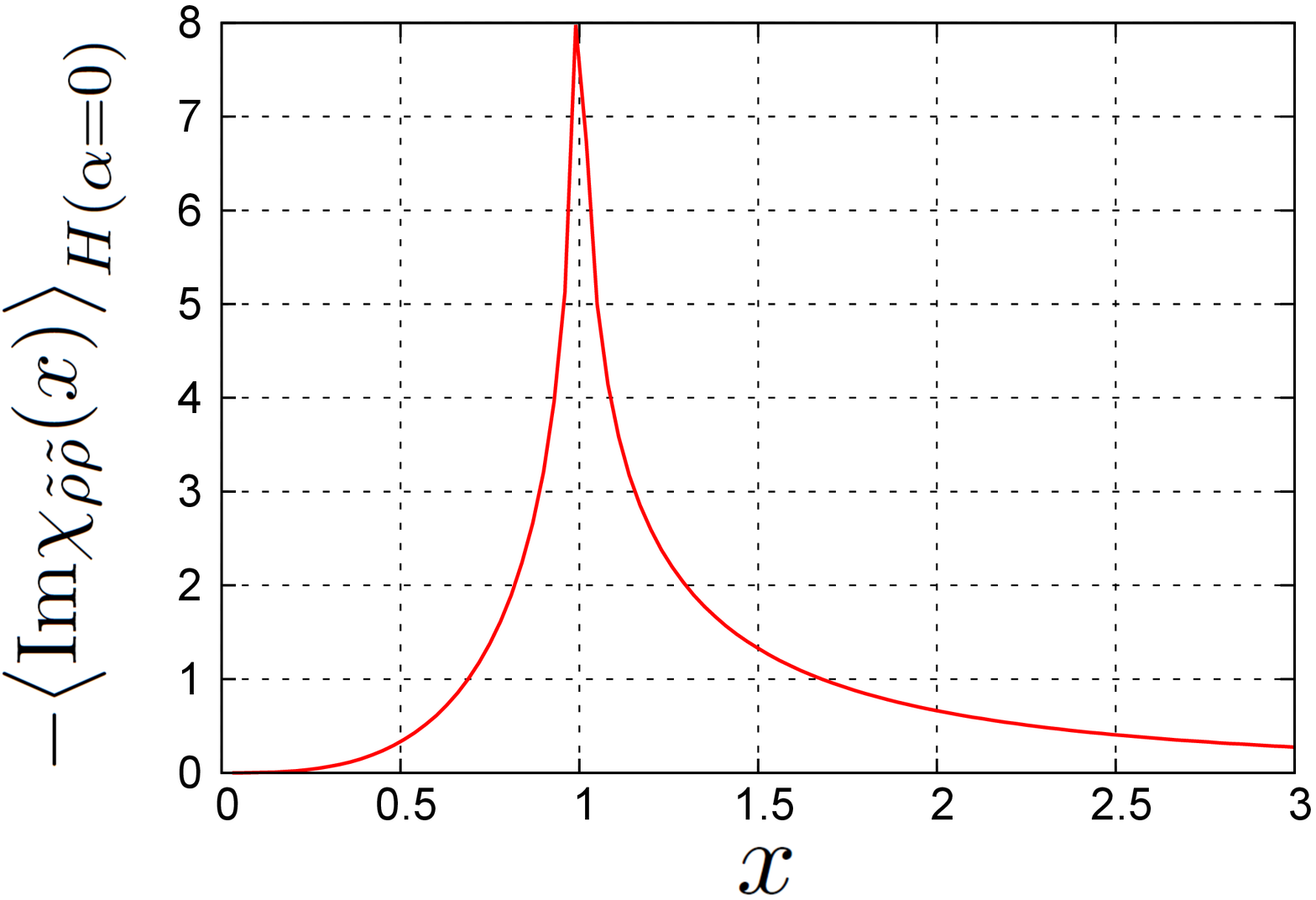}
\hspace{10mm}
\includegraphics[width=7cm]{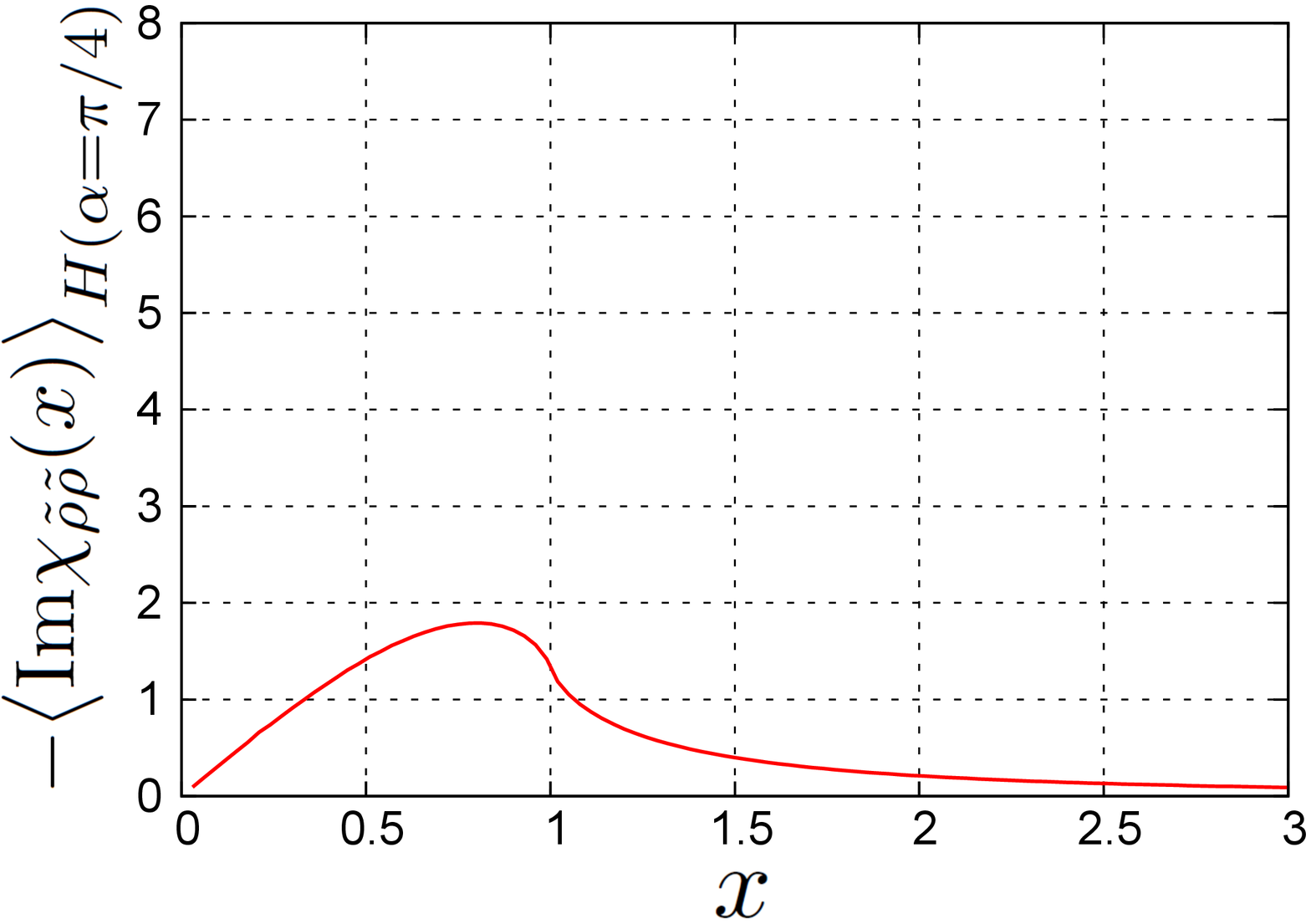}
\end{center}
\caption{
The electronic Raman scattering intensity for various Raman shifts $x=\omega / 2 \Delta_{0}$
at the field angle $\alpha = 0, \pi/4$.
}
\label{f4}
\end{figure}
%
%
%
%

%
%
%
%
\begin{figure}[htbp]
\begin{center}
\includegraphics[width=7cm]{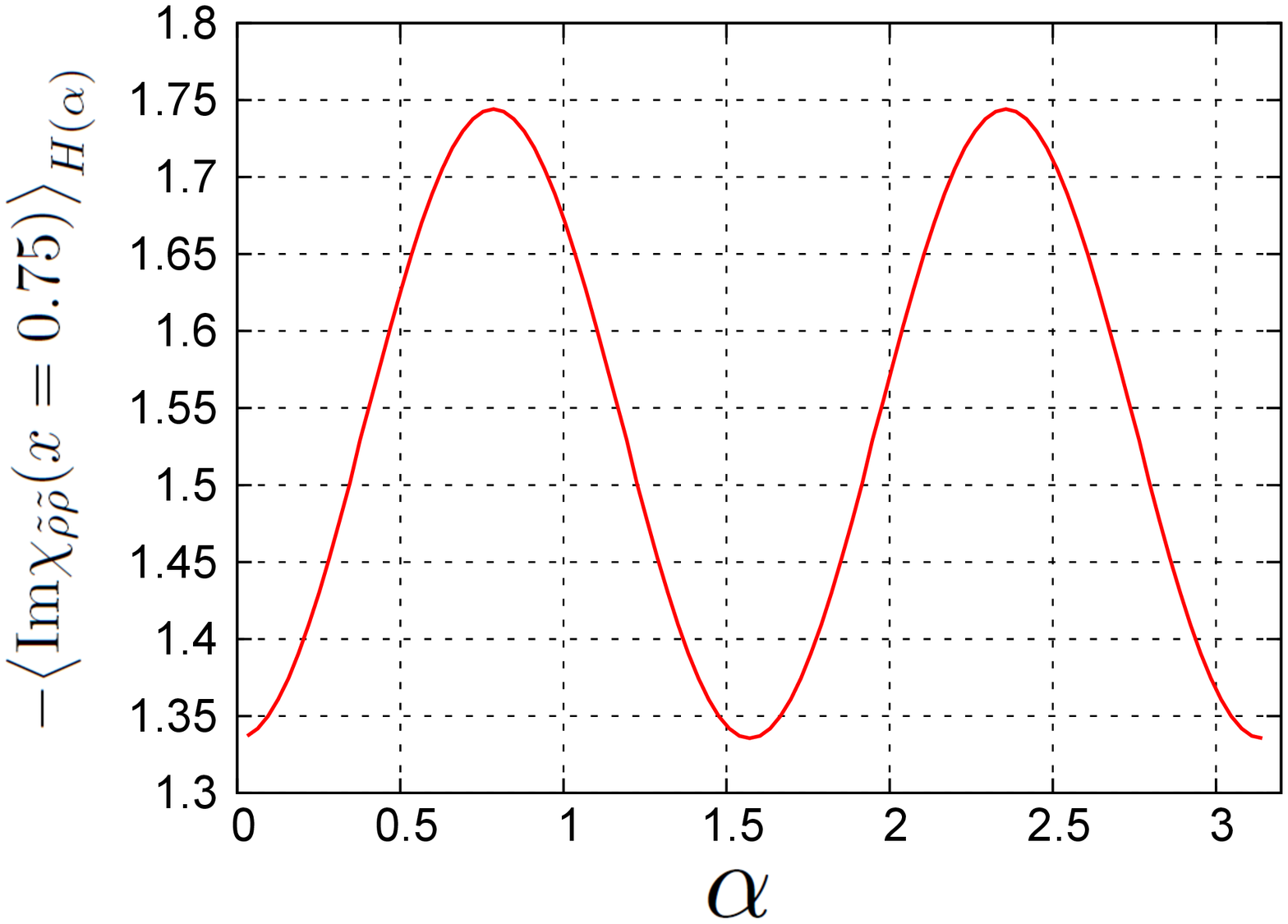}
\hspace{10mm}
\includegraphics[width=7cm]{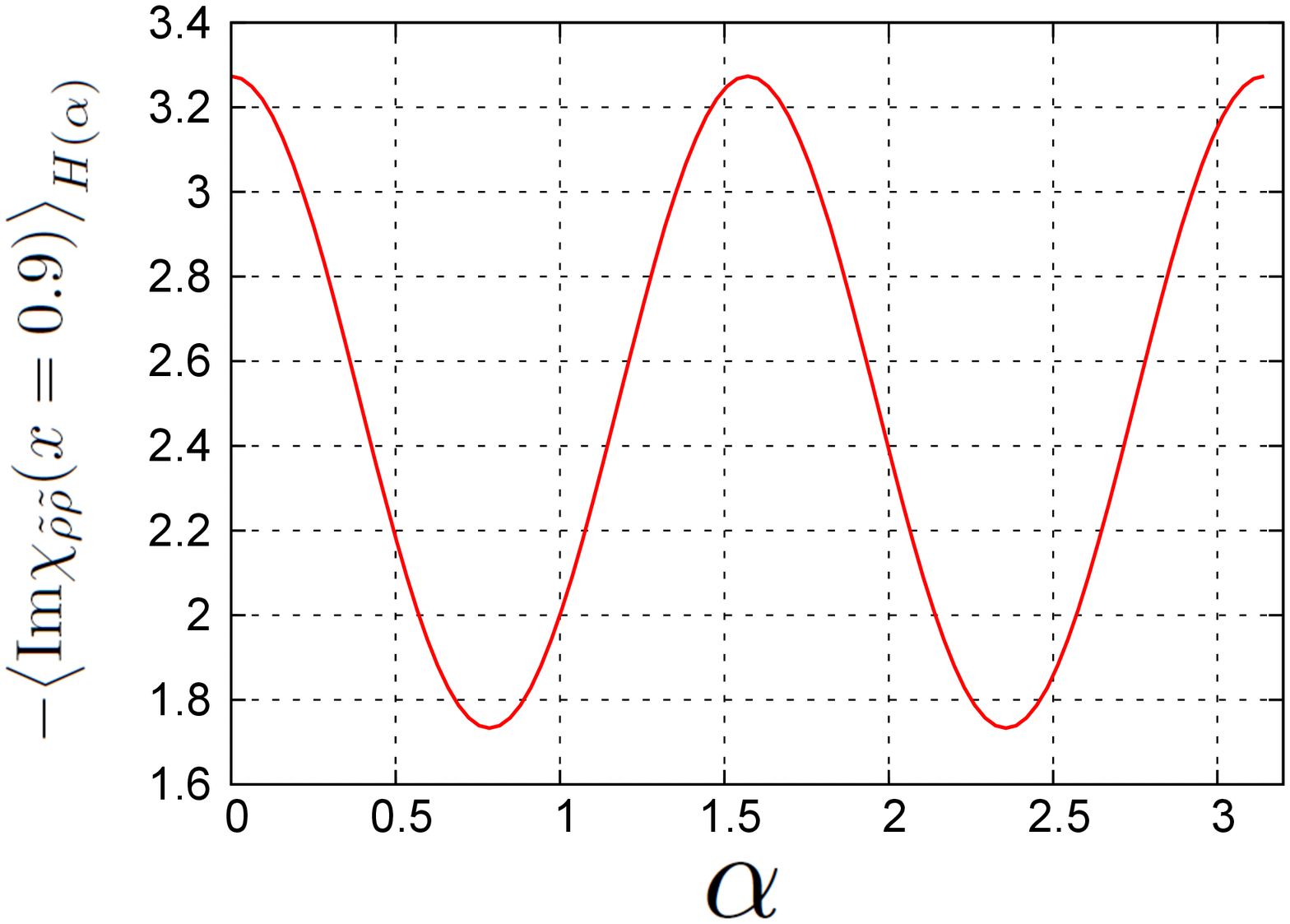}
\end{center}
\caption{
The electronic Raman scattering intensity for the various field angle $\alpha$
at the Raman shift energy $x=0.75,0.9$.
}
\label{f5}
\end{figure}
%
%
%
%

The electronic Raman scattering intensities for various Raman energy shifts $\omega$ and field angles $\alpha$
are plotted in Fig. \ref{f3}.
The left figure shows a plot of the electronic Raman scattering intensity function
$- \langle {\rm Im} \chi_{\tilde{\rho} \tilde{\rho}}(x) \rangle_{H(\alpha)}$
for various field angles $\alpha$ and normalized Raman shift $x=\omega / 2 \Delta_{0}$.
The right panel shows a top-down view of the 3-dimensional plot.

The results when a magnetic field is applied in the anti-nodal direction ($\alpha = 0$) and nodal direction ($\alpha = \pi/4$) are plotted in Fig. \ref{f4}.
When the field angle $\alpha$ is 0,
the intensity has logarithmic divergence at $x=1$ ($\omega = 2 \Delta_{0}$).
In the absence of an applied magnetic field, the electronic Raman scattering intensity always diverges for $d_{x^{2} - y^{2}}$ pairing symmetry and $B_{1g}$ polarization.
However, with the field angle $\alpha = \pi/4$,
the intensity does not diverge for any $\omega > 0$.

A novel phenomenon,
the reversal of extrema for various $\alpha$ with changing $\omega$,
is indicated in Fig. \ref{f5}.
When $x < 0.75$,
the $\alpha$-dependent intensity
has maxima (minima) at
$\alpha = \pi/4, 3 \pi/4$
\
$( \alpha = 0, \pi/2, \pi )$.
In contrast,
when $x>0.9$,
the intensity
has maxima (minima) at
$ \alpha = 0, \pi/2, \pi$
\
$( \alpha = \pi/4, 3 \pi/4 )$.
The phase shifts by $\pi/4$ between $x=0.75$ and $x=0.9$.

\section{Conclusion}

In contrast to the usual field-angle $\alpha$ resolved experiments,
energy (Raman shift energy $\omega$) is the essential variable for electronic Raman scattering experiments.
We find the novel phenomenon that
extrema of the electronic Raman scattering intensity can reverse
as a function of the Raman shift energy for constant temperature and field magnitude.

The present method may be applied to other superconducting symmetries,
such as $p$-wave or noncentrosymmetric superconductors
\cite{Klam2009}.
Application to other polarizations, such as $A_{1g}$ or $B_{2g}$, is also possible.

${}$

We thank Professor Kazuo Ueda for valuable discussions.

${}$

Professor Hayashi died after the time that the essential part of this study was conducted.
I wish to express my gratitude for his guidance and mentoring and offer my condolences to his family and friends.



\end{document}